\documentclass[11pt]{article}
\usepackage{amssymb}
\usepackage{latexsym}
\begin{document}
\setlength{\baselineskip}{15pt}
\title{ \mbox{} \vspace{-1.5cm} \\
		New global solutions of the 
		\\ Jacobi partial differential equations  
		\\ \mbox{}}
\author{\Large Benito Hern\'{a}ndez-Bermejo$^*$}
\date{}

\maketitle

\vspace{-1cm}
\begin{center}
{\em Departamento de F\'{\i}sica. Universidad Rey Juan Carlos. \\
Edificio Departamental II. Calle Tulip\'{a}n S/N. 28933--M\'{o}stoles--Madrid. Spain. }
\end{center}

\mbox{}

\mbox{}

\noindent \rule{15.6cm}{0.01in}

\vspace{2mm}

\noindent {\bf Abstract} \vspace{2mm}

A new family of solutions of the Jacobi partial differential equations for finite-dimensional Poisson systems is investigated. Such family is mathematically remarkable, as far as the functional dependences of the solutions appear to be associated to the distinguished invariants of the solutions themselves. This kind of Poisson structures (termed distinguished solutions or D-solutions) are defined for every nontrivial combination of values of the dimension and the rank, and are also determined in terms of functions of arbitrary nonlinearity, properties usually not present simultaneously in the already known solution families. In addition, D-solutions display several properties allowing the generation of an infinity of D-solutions from a given one, which is an uncommon feature in the framework of the Jacobi equations. Furthermore, a special family of D-solutions complying to the previous requirements is constructively characterized and analyzed. Examples are discussed focusing on physical implications and including an application for the global construction of the Darboux canonical form.

\mbox{}

\noindent {\em PACS:} 02.30.Jr, 45.20.-d, 45.20.Jj.


\mbox{}

\noindent {\em MSC:} 35Q72, 37J15, 37N05. 


\vspace{3mm}

\noindent {\em Keywords:} Jacobi partial differential equations; Finite-dimensional Poisson systems; Poisson structures; Casimir invariants.

\vspace{2mm}

\noindent \rule{15.6cm}{0.01in}

\mbox{}

\vfill

\mbox{}

\footnoterule

$^*$ Phone: (+ 00 34) 914 88 73 91. \mbox{} Fax: (+ 00 34) 914 88 73 38. 

\hspace{1.5mm} {\em E-mail address:\/} benito.hernandez@urjc.es

\pagebreak

\begin{flushleft}
{\bf 1. Introduction}
\end{flushleft}

In the last decades there has been a constant interest in the search and classification of solutions of the Jacobi partial differential equations (PDEs in what is to follow). For example, see 
\cite{agfs,agz1},\cite{pad1}-\cite{pad3},\cite{gyn1}-\cite{byv1},\cite{byv4,auo1,olv1,pla1,tpjm} and references therein. Given a domain $\Omega \subset \mathbb{R}^n$ and a system of local coordinates $x \equiv (x_1, \ldots ,x_n)$ defined on it, the Jacobi PDEs are given by 
\begin{equation}
     \label{n1jjac}
     \sum_{l=1}^n ( J_{li} \partial_l J_{jk} + J_{lj} \partial_l J_{ki} + 
     J_{lk} \partial_l J_{ij} ) = 0 \:\; , \;\:\;\: i,j,k=1, \ldots ,n
\end{equation}
plus the additional skew-symmetry conditions:
\begin{equation}
     \label{n1jsksym}
     J_{ij} =  - J_{ji} \:\; , \;\:\;\: i,j=1, \ldots ,n
\end{equation}
Together, equations (\ref{n1jjac}) and (\ref{n1jsksym}) are often referred to as the Jacobi PDEs. In expression (\ref{n1jjac}) and in what follows, it is $ \partial_l \equiv \partial / \partial x_l$. For the sake of clarity, the tensor index notation sometimes used in this context will be avoided, since its use does not provide any particular advantage for the analysis to come. The $C^{\infty}( \Omega )$ structure functions $J_{ij}(x)$ are the entries of an $n \times n$ structure matrix ${\cal J}$ which can be degenerate in rank. In this context, it is of paramount interest the characterization of $n$-dimensional solution families (namely $n \times n$ structure matrices) with both $n$ and the solution rank arbitrary, and being defined in terms of functions of arbitrary nonlinearity. The reason is that while structure matrices of lower dimension ($n=3$ and also $n=4$), lower rank (rank 2 structures, typically) and lower nonlinearity (specially linear, affine-linear, quadratic and cubic) have been widely investigated \cite{agz1,7},\cite{gyn1}-\cite{bs4},\cite{byv1},\cite{17,pla1,27}, the number of known families of arbitrary dimension (just to cite one of the previous three requirements) is very limited. In this sense, see for instance 
\cite{pad1}-\cite{pad3},\cite{bs07,byv4,auo1,olv1,pla1} and the discussions included in such works. 

The primary reason underlying the interest of the Jacobi PDEs is their key role in the construction of Poisson systems. Mathematically, a finite-dimensional dynamical system is said to be a Poisson system or to admit a Poisson structure \cite{lic1,olv1,wei1} if it can be written in terms of a set of smooth ODEs of the form: 
\begin{equation}
    \label{n1nham}
    \frac{\mbox{d}x_i}{{\mbox{d}t}} = \sum_{j=1}^n J_{ij}(x) \partial _j H(x) 
	\; , \;\:\; i = 1, \ldots , n, 
\end{equation} 
or briefly $\dot{x}= {\cal J}(x) \cdot \nabla H(x)$, where ${\cal J} \equiv (J_{ij})$ is a structure matrix and the smooth function $H$, which is by construction a time-independent first integral, plays the role of Hamiltonian. For instance, see \cite{olv1} and references therein for an overview and a historical discussion on Poisson systems. There are several reasons justifying the importance and flexibility of the Poisson representation. One is that finite-dimensional Poisson systems often arise in diverse fields of applied mathematics and physics, such as in dynamical systems theory \cite{5},\cite{pad1}-\cite{pad3},\cite{gyn1}, mathematical biology \cite{biyp,byv2,jmaa},\cite{pla1}-\cite{28}, mechanics 
\cite{pad1}-\cite{pad3},\cite{iib1,olv1,sim1}, electromagnetism \cite{cyl1,7}, plasma physics 
\cite{25}, optics \cite{7,dht3,17}, fluid mechanics \cite{nmm1}, etc. In fact, recasting a given dynamical system as a Poisson system allows the obtainment of a wide range of information which may be in the form of perturbative solutions \cite{cyl1,iib1}, invariants \cite{byv3,yhbm}, bifurcation properties and characterization of chaotic behaviour \cite{dht3,25}, efficient numerical integration \cite{mac1}, integrability results \cite{biyp},\cite{pad1}-\cite{pad3},\cite{mag1,olv1}, reductions \cite{biyp,7}, as well as stability analysis in terms of either the energy-Casimir algorithm \cite{jmaa}-\cite{17} or the energy-momentum method \cite{sim1}, to cite a sample. Accordingly, it is clear that the issue of recasting a given vector field not explicitly written in the form (\ref{n1nham}) in terms of a Poisson structure is a basic question in this context, which still remains as an open problem (for instance, see the discussions in \cite{5,7,gyn1,byv1}). This is a nontrivial decomposition to which important efforts have been devoted in past years in a variety of approaches, which also explains the need of finding a suitable structure matrix for the vector field under consideration. Consequently, reporting solutions of the Jacobi identities (\ref{n1jjac}-\ref{n1jsksym}) is a fundamental problem in this context. 

Another clue for the success of Poisson systems is that such representation provides a wide generalization of classical Hamiltonian systems, not only allowing for odd-dimensional vector fields, but also because a structure matrix verifying (\ref{n1jjac}-\ref{n1jsksym}) admits a great diversity of forms apart from the classical symplectic matrix \cite{z1gol}. At the same time, the possible rank degeneracy of the structure matrix ${\cal J}$ implies that in a Poisson system certain smooth and Hamiltonian-independent first integrals can exist, these being denoted $D(x)$ in what follows, and termed Casimir invariants (or also distinguished functions) in the literature. There is no analog in the framework of Hamiltonian systems for such constants of motion, which have the property of being the solution set of the system of coupled PDEs: 
\[
{\cal J} \cdot \nabla D =0
\]
On the other hand, there is a dynamical equivalence (at least locally, in a neighborhood $\Omega$ of every point, provided Rank(${\cal J}$) is constant in $\Omega$) between Poisson systems and Hamiltonian systems, as stated by Darboux theorem \cite{olv1}. This justifies that Poisson systems can be regarded, to a large extent, as a natural generalization of Hamiltonian systems. 

The applied context outlined throughout the previous considerations, together with the intrinsic mathematical interest of the Jacobi PDEs (\ref{n1jjac}-\ref{n1jsksym}), explain the present-day attention deserved by the investigation of new solution families of structure matrices, specially when these have some of the general features mentioned (namely admitting arbitrary values of the dimension and the rank, as well as being defined in terms of functions of arbitrary nonlinearity). 

In this article a new $n$-dimensional family of solutions of the Jacobi equations 
(\ref{n1jjac}-\ref{n1jsksym}) is investigated. Such family is mathematically remarkable, as far as the functional dependences of the structure functions appear to be associated to those of the Casimir invariants of the structure matrix itself. In particular, this kind of solutions (termed distinguished solutions or D-solutions, as it will be detailed in brief) is defined for every possible nontrivial combination of values of the dimension and the rank, and is determined in terms of smooth functions of arbitrary nonlinearity. In addition, D-solutions display properties allowing the generation of an infinity of D-solutions from a given one. As it will be pointed out later, this feature is interesting as far as such kind of properties is quite uncommon (and usually of a very limited scope, when present \cite{bs1,bs2,byv2}) in the framework of the Jacobi equations (\ref{n1jjac}-\ref{n1jsksym}). Furthermore, some families of D-solutions of special interest, globally defined and complying to the previous requirements, are constructively characterized and analyzed. 

The structure of the article is the following. In Section 2 the definition and analysis of the new family of solutions of the Jacobi equations are presented. A discussion of such results, complemented with several instances of the theory, is provided in Section 3. A special family of three-dimensional D-solutions is characterized in Section 4, which leads to some physical and applied considerations. The work is concluded with some final remarks in Section 5.

\mbox{}

\mbox{}

\begin{flushleft}
{\bf 2. Distinguished Jacobi equations and distinguished Poisson structures}
\end{flushleft}

We begin with a description of the problem:

\mbox{}

\noindent {\bf Definition 1.} {\em Let ${\cal J} \equiv (J_{ij})$ be an $n \times n$ matrix 
defined in a domain (open and simply connected set) $\Omega \subset \mathbb{R}^n$ ($n \geq 3$) and composed by $C^{\infty}( \Omega )$ real functions $J_{ij}(x)$. Then ${\cal J}$ is said to be a solution of the distinguished Jacobi equations in $\Omega$ if for every $x \in \Omega$ it is skew-symmetric and
\begin{equation}
\label{n1rjac}
	\sum_{l=1}^n J_{kl} \partial _l J_{ij} = 0 \;\: , \;\:\;\: i,j,k = 1, \ldots , n
\end{equation}
or, equivalently, if ${\cal J}^T=-{\cal J}$ and ${\cal J}\cdot \nabla J_{ij}=0$ for all $i,j=1, \ldots ,n$, where the superscript $^T$ denotes the transpose matrix. Every matrix ${\cal J}$ being a solution of the distinguished Jacobi equations (\ref{n1jsksym},\ref{n1rjac}) will be termed a distinguished solution, or a D-solution.
}

\mbox{}

This definition implies two relevant consequences that can be summarized as follows: 

\mbox{}

\noindent {\bf Corollary 1.} {\em Let ${\cal J} \equiv (J_{ij})$ be a D-solution defined in a domain $\Omega \subset \mathbb{R}^n$, then:
\begin{description}
\item[\mbox{\rm \mbox{\em (a)}}] ${\cal J}$ is a solution of the Jacobi equations 
	(\ref{n1jjac}--\ref{n1jsksym}), and therefore it is a structure matrix in $\Omega$. The 	converse is obviously not true, namely not every structure matrix is a D-solution.
\item[\mbox{\rm \mbox{\em (b)}}] Functions $J_{ij}(x)$ are Casimir invariants of $\cal J$ globally defined in $\Omega$ for all $i,j=1, \ldots ,n$. 
\end{description}
}

\mbox{}

Therefore, briefly speaking D-solutions can be described as structure matrices defined in terms of their own Casimir invariants. As indicated, Casimir invariants are also termed ``distinguished functions'' in the literature on Poisson systems \cite{olv1}. This is the reason accounting for the denomination of ``distinguished solutions'' given here to the present kind of structure matrices entirely composed of Casimir invariants. In spite that D-solutions appear naturally as a nice special case of Poisson structures, it is remarkable that they have not been previously reported in the literature (to the author's knowledge). Later in this section, a wide family of D-solutions will be constructed and characterized in full detail. Before that, it is interesting to further focus on some general properties associated to 
D-solutions. Such properties are not present in general in Poisson structures, but can be efficiently determined in D-solutions. In order to see this, a preliminary definition is convenient: 

\mbox{}

\noindent {\bf Definition 2.} {\em Let ${\cal M} \equiv (M_{ij}(x))$ be an $n \times n$ real matrix defined in a domain $\Omega \subset \mathbb{R}^n$ and such that functions $M_{ij}(x)$ are $C^{\infty}(\Omega)$ for all $i,j=1, \ldots ,n$. A $C^{\infty}(\Omega)$ function $f(x): \Omega \rightarrow \mathbb{R}$ is said to be kernel-gradient (or KG) for matrix ${\cal M}$ if ${\cal M} \cdot \nabla f =0$ for every $x \in \Omega$.
}

\mbox{}

The previous definition is natural in this context, since a D-solution is just a skew-symmetric matrix for which all the entries belong to the set of its KG functions. This point of view will be useful in brief because a $C^{\infty}$ function $h(f_1, \ldots ,f_p)$ of one or more KG functions $f_1, \ldots ,f_p$ is also a KG function of the same matrix. We can now state a first result:

\mbox{}

\noindent {\bf Theorem 1.} {\em Let ${\cal J} \equiv (J_{ij})$ be an $n \times n$ D-solution defined in a domain $\Omega \subset \mathbb{R}^n$ and such that Rank($\cal J$) $=r$ in $\Omega$. Let $(D_{r+1}(x), \ldots ,D_n(x))$ be a complete set of functionally independent Casimir invariants of $\cal J$ in $\Omega$. In addition, let ${\cal A} \equiv (A_{ij})$ denote an $n \times n$ matrix of entries of the form $A_{ij}(D_{r+1}(x), \ldots , D_n(x))$ such that the real functions $A_{ij}(y_1, \ldots , y_{n-r})$ are $C^{\infty}(\mathbb{R}^{n-r})$ for all $i,j=1, \ldots ,n$. Then:
\begin{description}
\item[\mbox{\rm \mbox{\em (a)}}] For every odd real polynomial of ${\cal J}$, 
\begin{equation}
\label{n1podd}
	P({\cal J}) = \sum_{i=0}^k a_{2i+1} {\cal J}^{2i+1} \:\; , \;\:\;\: k \geq 0,
\end{equation}
and for every skew-symmetric matrix ${\cal A}$ of the aforementioned kind, the matrix product $(P({\cal J}) \cdot {\cal A})^m \cdot P({\cal J})$ is a D-solution in $\Omega$ for every integer $m \geq 0$.
\item[\mbox{\rm \mbox{\em (b)}}] For every odd real polynomial $P({\cal J})$ of ${\cal J}$ of the form (\ref{n1podd}) and for every symmetric matrix ${\cal A}$ of the kind indicated which commutes with $\cal J$, the product $({\cal A} \cdot P({\cal J}))^m$ is a D-solution in $\Omega$ for every odd integer $m \geq 1$.
\item[\mbox{\rm \mbox{\em (c)}}] For every odd real polynomial $P({\cal J})$ of ${\cal J}$ of the form (\ref{n1podd}) and for every matrix ${\cal A}$ of the kind indicated which commutes with $\cal J$, the product ${\cal A} \cdot P({\cal J}) \cdot {\cal A}^T$ is a D-solution in $\Omega$.
\item[\mbox{\rm \mbox{\em (d)}}] For every even real polynomial without constant term of ${\cal J}$, 
\[
	Q({\cal J}) = \sum_{i=1}^k a_{2i} {\cal J}^{2i} \:\; , \;\:\;\: k \geq 1,
\]
and for every skew-symmetric matrix ${\cal A}$ of the kind already indicated, the matrix product $(Q({\cal J}) \cdot {\cal A})^m \cdot Q({\cal J})$ is a D-solution in $\Omega$ for every odd integer $m \geq 1$.
\item[\mbox{\rm \mbox{\em (e)}}] If $\eta (y_1, \ldots ,y_{n-r}) : \mathbb{R}^{n-r} \rightarrow \mathbb{R}$ is an arbitrary $C^{\infty}(\mathbb{R}^{n-r})$ real function, then the product $ \eta (D_{r+1}(x), \ldots ,D_n(x)) {\cal J}$ is a D-solution in $\Omega$.
\end{description}
}

\mbox{}

\noindent {\bf Proof.} (a) All Casimir invariants of ${\cal J}$ are also KG functions for 
$(P({\cal J}) \cdot {\cal A})^m \cdot P({\cal J})$. Then, the form of ${\cal A}$ implies that every entry of $(P({\cal J}) \cdot {\cal A})^m \cdot P({\cal J})$ is a KG function of the matrix itself. To conclude, notice that $P({\cal J})$ is skew-symmetric:
\[
	[P({\cal J})]^T = \sum_{i=0}^k a_{2i+1} (-1)^{2i+1} {\cal J}^{2i+1} = - P({\cal J})
\]
According to this, $(P({\cal J}) \cdot {\cal A})^m \cdot P({\cal J})$ is also skew-symmetric:
\[
	[(P({\cal J}) \cdot {\cal A})^m \cdot P({\cal J})]^T = 
	- P({\cal J}) \cdot [{\cal A}^T \cdot (P({\cal J}))^T]^m = 
	- (P({\cal J}) \cdot {\cal A})^m \cdot P({\cal J})
\]

(b) The Casimir invariants of ${\cal J}$ are KG functions of $({\cal A} \cdot P({\cal J}))^m$. Due to the functional dependence of ${\cal A}$, matrix $({\cal A} \cdot P({\cal J}))^m$ is entirely composed by KG functions of $({\cal A} \cdot P({\cal J}))^m$. In addition, it can be seen that $({\cal A} \cdot P({\cal J}))^m$ is skew-symmetric because $P({\cal J})$ is, since:
\[
	[({\cal A} \cdot P({\cal J}))^m]^T = ([P({\cal J})]^T \cdot {\cal A}^T)^m = 
	(-P({\cal J}) \cdot {\cal A})^m = -({\cal A} \cdot P({\cal J}))^m
\]

(c) Matrix ${\cal A} \cdot P({\cal J}) \cdot {\cal A}^T$ is composed by Casimir invariants of ${\cal J}$. Moreover, ${\cal A} \cdot P({\cal J}) \cdot {\cal A}^T$ is skew-symmetric because $P({\cal J})$ is odd. In addition, it is simple to show that ${\cal J}$ and ${\cal A}^T$ commute if and only if ${\cal J}$ and ${\cal A}$ commute. In fact: 
\[
	{\cal J} \cdot {\cal A}^T - {\cal A}^T \cdot {\cal J} = 
	-{\cal J}^T \cdot {\cal A}^T + {\cal A}^T \cdot {\cal J}^T = 
	({\cal J} \cdot {\cal A} - {\cal A} \cdot {\cal J})^T = 0
\]
Then, $P({\cal J})$ also commutes with ${\cal A}$ and ${\cal A}^T$. For every $i,j=1, \ldots ,n$, let us denote as $\tilde{J}_{ij}$ the entries of ${\cal A} \cdot P({\cal J}) \cdot {\cal A}^T$. We then have: 
\[
	{\cal A} \cdot P({\cal J}) \cdot {\cal A}^T \cdot \nabla \tilde{J}_{ij} = 
	{\cal A} \cdot {\cal A}^T \cdot P({\cal J}) \cdot \nabla \tilde{J}_{ij} = 0
\]
The last step holds because $\tilde{J}_{ij}$ is a Casimir invariant of ${\cal J}$. Consequently, $\tilde{J}_{ij}$ is a KG function of ${\cal A} \cdot P({\cal J}) \cdot {\cal A}^T$ for all $i,j$, and it is proved that such matrix is a D-solution. 

\mbox{}

(d) Every Casimir invariant of ${\cal J}$ is a KG function of $(Q({\cal J}) \cdot {\cal A})^m \cdot Q({\cal J})$, so that taking into account the form of ${\cal A}$, we have that $(Q({\cal J}) \cdot {\cal A})^m \cdot Q({\cal J})$ is composed by its own KG functions. The proof will be complete after verifying the skew-symmetry of $(Q({\cal J}) \cdot {\cal A})^m \cdot Q({\cal J})$. To see this, note first that $Q({\cal J})$ is symmetric:
\[
	[Q({\cal J})]^T = \sum_{i=1}^k a_{2i} (-1)^{2i} {\cal J}^{2i} = Q({\cal J})
\]
And consequently, we have: 
\[ 
	[(Q({\cal J}) \cdot {\cal A})^m \cdot Q({\cal J})]^T = 
	Q({\cal J}) \cdot [-{\cal A} \cdot Q({\cal J})]^m = 
	- (Q({\cal J}) \cdot A)^m \cdot Q({\cal J}) 
\]

(e) Matrix $\eta (D_{r+1}(x), \ldots ,D_n(x)) {\cal J}$ is obviously skew-symmetric. The Casimir invariants of ${\cal J}$ are KG functions of $\eta {\cal J}$, and consequently this matrix is fully composed of its own KG functions. 

This completes the proof of the theorem. \hfill $\Box$

\mbox{}

In the previous theorem, it is worth noting that no assumptions are being made on the rank of ${\cal A}$, in contrast with the situation to be reported in Theorem 2. Note also that, in particular, matrix ${\cal A}$ can be constant. Regarding statement (e) of Theorem 1, it is interesting to note that the multiplicative deformations preserving a Poisson structure have been analyzed in detail in the literature \cite{bnn2,bnn4}. Actually, from these references it is known that every expression of the form $\eta (D_{r+1}(x), \ldots ,D_n(x)) {\cal J}(x)$ remains a structure matrix provided ${\cal J}(x)$ is. However, what is relevant in the present context is that such kind of operations also preserves the property of being a D-solution. 

As indicated, the type of properties described in Theorem 1, when present in the context of the Jacobi PDEs, deserves some interest for several reasons. First, because Theorem 1 is not limited from the point of view of the dimension or the rank of the solutions to which it can be applied, and consequently it is remarkably general. And second, because such kind of properties is uncommon in the context of finite-dimensional Poisson structures. We see however that these properties are present in the specific domain of the distinguished problem. Some less general instances of this kind of properties for nondistinguished Poisson structures can be found in \cite{bs1,bs2} for certain three-dimensional structures and in \cite{byv2} for some $n$-dimensional solution families. 

The following result deals with variable transformations keeping invariant the property of being a D-solution: 

\mbox{}

\noindent {\bf Theorem 2.} {\em Let ${\cal J}(x) \equiv (J_{ij}(x))$ be an $n \times n$ D-solution defined in a domain $\Omega \subset \mathbb{R}^n$ and such that Rank($\cal J$) $=r$ in $\Omega$. Let $(D_{r+1}(x), \ldots ,D_n(x))$ be a complete set of functionally independent Casimir invariants of $\cal J$ in $\Omega$. In addition, let $y= \phi (x)$ be a change of variables which is $C^{\infty}(\Omega)$, diffeomorphic everywhere in $\Omega$ and such that its Jacobian is of the form $(\partial y_i/ \partial x_j) \equiv A_{ij}(D_{r+1}(x), \ldots , D_n(x))$ for all $i,j=1, \ldots ,n$. Then, after the application of the change of coordinates $y= \phi (x)$ to ${\cal J}(x)$, the resulting structure matrix ${\cal J}^*(y)$ is also a D-solution. 
}

\mbox{}

\noindent {\bf Proof.} Let ${\cal A}$ be the Jacobian of the transformation $y= \phi (x)$, namely:
\[
	{\cal A} = \frac{\partial (y_1, \ldots ,y_n)}{\partial (x_1, \ldots ,x_n)}
\]
Then, according to the chain rule, it can be seen that for every differentiable function $f(x)$ it is $\nabla _y f(y) = ({\cal A}^{-1})^T \cdot \nabla _x f(y(x))$, where the subindex in the nabla operator indicates the differentiation variables. In addition, it is worth recalling the coordinate transformation rule for structure matrices, which amounts to ${\cal J}^*(y) = {\cal A}(x(y)) \cdot {\cal J}(x(y)) \cdot {\cal A}^T(x(y))$. In particular, it is well-known that such ${\cal J}^*(y)$ is always a structure matrix. According to the hypotheses assumed we see that, by construction, all the entries of ${\cal J}^*(y)$ are Casimir invariants of ${\cal J}(x)$. Therefore, in order to complete the proof it is necessary to show that the entries $J^*_{ij}(y)$ of ${\cal J}^*(y)$ are Casimir invariants of ${\cal J}^*(y)$ for every $i,j=1, \ldots ,n$. For convenience let us define $\alpha_{ij}(x) \equiv A_{ij}(D_{r+1}(x), \ldots , D_n(x))$ for all $i,j=1, \ldots ,n$. We can now compute: 
\[
	\begin{array}{c}
	{\cal J}^*(y) \cdot \nabla_y J^*_{ij}(y) = {\cal A}(x) \cdot {\cal J}(x) \cdot 
	{\cal A}^T(x) \cdot ({\cal A}^{-1})^T(x) \nabla_x J^*_{ij}(y(x)) = \vspace{2mm} \\ 
	{\cal A}(x) \cdot {\cal J}(x) \cdot \mbox{\Large (}
	\sum_{k,l=1}^n \alpha_{ik}(x) J_{kl}(x) \nabla_x \alpha_{jl}(x) + \alpha_{ik}(x) 
	\alpha_{jl}(x) \nabla_x J_{kl}(x) 
	\vspace{2mm} \\ 
	+ \alpha_{jl}(x) J_{kl}(x) \nabla_x \alpha_{ik}(x) \mbox{\Large )} = 0 
	\end{array}
\]
The outcome amounts to zero because every $J_{kl}(x)$ is a Casimir invariant of ${\cal J}(x)$, and the same is true for all the entries $\alpha_{ij}(x) = A_{ij}(D_{r+1}(x), \ldots , D_n(x))$ of the Jacobian. Theorem 2 is thus proved. \hfill $\Box$

\mbox{}

Since constants are Casimir invariants for every structure matrix, an interesting outcome arising from Theorem 2 is the following:

\mbox{}

\noindent {\bf Corollary 2.} {\em The property of being a D-solution is preserved after every linear and invertible change of variables.}

\mbox{}

D-solutions also have some additional properties that are worth noting. In order to present them, we need a previous definition: 

\mbox{}

\noindent {\bf Definition 3.} {\em For every D-solution ${\cal J}$ defined in a domain $\Omega \subset \mathbb{R}^n$, the set of all the Casimir invariants of ${\cal J}$ in $\Omega$ is termed $\Delta [{\cal J}]$. 
}

\mbox{}

We can now state: 

\mbox{}

\noindent {\bf Theorem 3.} {\em Let ${\cal J}_1,\ldots,{\cal J}_p$ be a set of $n \times n$ D-solutions of constant rank $r \geq 2$ defined in a domain $\Omega \subset \mathbb{R}^n$ and such that $\Delta[{\cal J}_1] = \ldots = \Delta[{\cal J}_p]$. Then: 
\begin{itemize}
\item[\mbox{\rm \mbox{\em (a)}}] In the case $p=2$, the matrix 
\begin{equation}
\label{prcg1}
	{\cal J}_+= {\cal J}_1+{\cal J}_2
\end{equation}
is a D-solution defined in $\Omega$ such that $\Delta[{\cal J}_i] \subseteq \Delta[{\cal J}_+]$ for $i=1,2$. 
\item[\mbox{\rm \mbox{\em (b)}}] For every $p \geq 1$, the matrix
\begin{equation}
\label{prcg2}
	{\cal J}_{\perp}= \prod_{i=1}^p {\cal J}_i + (-1)^{p+1} \prod_{i=1}^p {\cal J}_{p-i+1} = 
	{\cal J}_1 \cdot \ldots \cdot {\cal J}_p + (-1)^{p+1}{\cal J}_p \cdot \ldots \cdot 
	{\cal J}_1
\end{equation}
is a D-solution defined in $\Omega$ such that $\Delta[{\cal J}_i] \subseteq \Delta[{\cal J}_{\perp}]$ for $i=1, \ldots ,p$. 
\end{itemize}
}

\mbox{}

\noindent {\bf Proof.} (a) It is clear that ${\cal J}_+$ is skew-symmetric and that every entry of ${\cal J}_+$ in (\ref{prcg1}) is a KG function of such matrix. The proof can be completed with suitable examples. If we choose ${\cal J}_2 = -{\cal J}_1$, we obtain $\Delta[{\cal J}_1]=\Delta[{\cal J}_2] \subset \Delta[{\cal J}_+]$. In addition, if ${\cal J}_2 = {\cal J}_1$, we have $\Delta[{\cal J}_1]=\Delta[{\cal J}_2] = \Delta[{\cal J}_+]$. We thus see that $\Delta[{\cal J}_i] \subseteq \Delta[{\cal J}_+]$ for $i=1,2$, and the proof of statement (a) is complete.

\mbox{}

(b) Let us first show that ${\cal J}_{\perp}$ is skew-symmetric: 
\[
	\begin{array}{c} 
	{\cal J}_{\perp}^T = {\cal J}_p^T \cdot \ldots \cdot {\cal J}_1^T + 
				(-1)^{p+1}{\cal J}_1^T \cdot \ldots \cdot {\cal J}_p^T =  
				\vspace{2mm} \\ 
			   (-1)^p{\cal J}_p \cdot \ldots \cdot {\cal J}_1 + 
				(-1)^{2p+1}{\cal J}_1 \cdot \ldots \cdot {\cal J}_p = -{\cal J}_{\perp} 
	\end{array}
\]
Since by hypothesis we have $\Delta[{\cal J}_1] = \ldots = \Delta[{\cal J}_p]$, it is clear that ${\cal J}_{\perp}$ is by construction composed by Casimir invariants of the ${\cal J}_i$, which in turn must be KG functions of ${\cal J}_{\perp}$. Therefore ${\cal J}_{\perp}$ is a D-solution. As before, the proof can now be completed by means of examples. For instance, let us consider the case $p=2$ with ${\cal J}_1={\cal J}_2$, then it is ${\cal J}_{\perp}=O_{n \times n}$, where the symbol $O$ denotes the null matrix of size indicated by the respective subindex. We then have $\Delta[{\cal J}_1] = \Delta[{\cal J}_2] \subset \Delta[{\cal J}_{\perp}]$. Alternatively, now let $p=1$, then it is ${\cal J}_{\perp}=2{\cal J}_1$ and clearly $\Delta[{\cal J}_1] = \Delta[{\cal J}_{\perp}]$. Then we see that $\Delta[{\cal J}_i] \subseteq \Delta[{\cal J}_{\perp}]$ for $i=1, \ldots , p$, and this completes the proof of Theorem 3. \hfill $\Box$

\mbox{}

In particular, note that operation (\ref{prcg2}) is a generalization of the commutator (case $p=2$). 

Together, Theorems 1, 2 and 3 show that D-solutions display a unusually rich framework of transformation properties for Poisson structures. It is worth stressing that such a situation is remarkable in the context of Jacobi PDEs. Clearly, these results suggest that D-solutions constitute a solution family previously unknown but of special interest. 

The fact that D-solutions are given in terms of their Casimir invariants is mathematically appealing but, at the same time, the issue of the practical determination of some representative D-solutions requires further investigation. This is actually possible: as anticipated, after the previous definitions and general properties the aim now is to consider one family of D-solutions which is amenable to constructive characterization. This is the content of the next result:

\mbox{}

\noindent {\bf Theorem 4.} {\em Let $n \geq 3$ and $\rho \leq n$ be two positive integers, and consider the ($n- \rho $) functions 
\begin{equation}
\label{n1casl}
	D_l(x) = x_l - \sum_{k=1}^{\rho} a_{lk}x_k \;\:\; , \:\;\:\;\;\: l = \rho + 1, \ldots ,n
\end{equation}
where $a_{lk}$ are real constants for all $l,k$. In addition, for $i,j=1, \ldots , \rho$, let $\psi_{ij}(y_1, \ldots ,y_{n- \rho})$ be $C^{\infty}(\mathbb{R}^{n- \rho})$ functions that are 
skew-symmetric in their subindexes, namely $\psi_{ij}(y_1, \ldots ,y_{n- \rho}) = - \psi_{ji}(y_1, \ldots ,y_{n- \rho})$ for all $i,j$. Finally, let ${\cal J} \equiv (J_{ij})$ be the $n \times n$ matrix given by:
\begin{equation}
\label{n1jlin}
	J_{ij}(x) = \left\{
	\begin{array}{lcl}
	\psi_{ij}(D_{\rho +1}(x), \ldots ,D_n(x)) & , & i,j=1, \ldots , \rho  \\
	\displaystyle{ \sum_{k=1}^{\rho} a_{jk} \psi_{ik}(D_{\rho +1}(x), \ldots ,
				D_n(x)) }& , & i=1, \ldots , \rho \; ; \; j= \rho + 1, \ldots , n \\
	\displaystyle{ \sum_{k=1}^{\rho} a_{ik} \psi_{kj}(D_{\rho +1}(x), \ldots ,
				D_n(x)) }& , & i= \rho + 1, \ldots , n \; ; \; j=1, \ldots , \rho   \\
	\displaystyle{ \sum_{k,l=1}^{\rho} a_{ik} a_{jl} \psi_{kl}(D_{\rho +1}(x), 
				\ldots , D_n(x)) }& , & i,j= \rho + 1, \ldots , n 
	\end{array} \right.
\end{equation}
Then, ${\cal J}$ is a D-solution of the Jacobi equations which is globally defined in $\mathbb{R}^n$ and such that Rank(${\cal J}$)$\leq \rho - \rho \bmod 2$ for every $x \in \mathbb{R}^n$. In addition, the ($n- \rho$) functions $D_l(x)$ in (\ref{n1casl}) constitute everywhere in $\mathbb{R}^n$ a set of functionally independent Casimir invariants of ${\cal J}$.
}

\mbox{}

\noindent {\bf Proof.} The proof is constructive. For this, let us first consider the submatrix structure of ${\cal J}$ as suggested by equation (\ref{n1jlin}), namely:
\begin{equation}
\label{n1subj}
	{\cal J} \equiv \left( \begin{array}{ccc} 
		{\cal J}^{[1]} & \vline & {\cal J}^{[2]} \\ \hline
		{\cal J}^{[3]} & \vline & {\cal J}^{[4]}
	    \end{array} \right)
\end{equation}
where ${\cal J}^{[1]}$, ${\cal J}^{[2]}$, ${\cal J}^{[3]}$ and ${\cal J}^{[4]}$ are submatrices of sizes $\rho \times \rho$, $\rho \times (n- \rho)$, $(n- \rho) \times \rho$ and $(n - \rho) \times (n- \rho)$, respectively. In the rest of the proof, the entries of ${\cal J}^{[k]}$ will be denoted $J_{ij}^{[k]}$ for all $k=1, \ldots ,4$. Therefore, according to (\ref{n1jlin}) and (\ref{n1subj}) we have $J_{ij}^{[1]}=\psi_{ij}(D_{\rho +1}(x), \ldots ,D_n(x))$ for all $i,j = 1, \ldots , \rho$. Regarding ${\cal J}^{[2]}$, note that from (\ref{n1jlin}) we have:
\begin{equation}
\label{n1j2}
	J_{ij}^{[2]} = \sum_{k=1}^{\rho} a_{jk} J_{ik}^{[1]} 
	\;\: , \;\:\;\: i =1, \ldots , \rho \;\: , \;\: j = \rho +1, \ldots ,n
\end{equation}
Similarly for ${\cal J}^{[3]}$, equation (\ref{n1jlin}) implies that:
\begin{equation}
\label{n1j3}
	J_{ij}^{[3]} = \sum_{k=1}^{\rho} a_{ik} J_{kj}^{[1]} 
	\;\: , \;\:\;\: i = \rho +1, \ldots ,n \;\: , \;\: j =1, \ldots , \rho
\end{equation}
To conclude, notice that from (\ref{n1jlin}) we also obtain for ${\cal J}^{[4]}$: 
\begin{equation}
\label{n1j41}
	J_{ij}^{[4]} = \sum_{k,l=1}^{\rho} a_{ik} a_{jl} J_{kl}^{[1]} 
	\;\: , \;\:\;\: i,j = \rho +1, \ldots ,n
\end{equation}
Taking (\ref{n1j2}) into account, we can express (\ref{n1j41}) as a result very similar to that for ${\cal J}^{[3]}$ in (\ref{n1j3}):
\begin{equation}
\label{n1j42}
	J_{ij}^{[4]} = \sum_{k=1}^{\rho} a_{ik} J_{kj}^{[2]} 
	\;\: , \;\:\;\: i,j = \rho +1, \ldots ,n
\end{equation}
For what is to come, it is also necessary to observe that for ${\cal J}^{[4]}$ there is another expression analogous to the one displayed in (\ref{n1j42}). Now such alternative dependence can be obtained from (\ref{n1j3}) and (\ref{n1j41}) in terms of ${\cal J}^{[3]}$ as:
\begin{equation}
\label{n1j43}
	J_{ij}^{[4]} = \sum_{k=1}^{\rho} a_{jk} J_{ik}^{[3]} 
	\;\: , \;\:\;\: i,j = \rho +1, \ldots ,n
\end{equation}
With the help of expressions (\ref{n1j2}--\ref{n1j43}) some auxiliary results can be provided now:

\mbox{}

\noindent {\bf Lemma 1.} {\em Matrix ${\cal J} \equiv (J_{ij})$ in (\ref{n1jlin}) is 
skew-symmetric.}

\mbox{}

\noindent {\bf Proof.} Submatrix ${\cal J}^{[1]}$ is skew-symmetric by definition. Let us now prove that submatrices ${\cal J}^{[2]}$ and ${\cal J}^{[3]}$ also verify skew-symmetry. According to (\ref{n1j2}) and (\ref{n1j3}) we have:
\[
	J_{ij}^{[2]} + J_{ji}^{[3]} = 
	\sum_{k=1}^{\rho} a_{jk} J_{ik}^{[1]} + \sum_{k=1}^{\rho} a_{jk} J_{ki}^{[1]} = 0 
	\;\: , \;\:\;\: i =1, \ldots , \rho \;\: , \;\: j = \rho +1, \ldots ,n
\]
To conclude, making use of (\ref{n1j41}) for ${\cal J}^{[4]}$ we have:
\[
	J_{ji}^{[4]} = \sum_{k,l=1}^{\rho} a_{jk} a_{il} J_{kl}^{[1]} = 
	- \sum_{k',l'=1}^{\rho} a_{ik'} a_{jl'} J_{k'l'}^{[1]} = - J_{ij}^{[4]}
	\;\: , \;\:\;\: i , j = \rho +1, \ldots ,n
\]
Lemma 1 is thus proved. \hfill $\Box$

\mbox{}

\noindent {\bf Lemma 2.} {\em Functions $D_l(x)$ in (\ref{n1casl}) are KG for matrix 
${\cal J}$ in (\ref{n1jlin}) for all $l= \rho +1, \ldots , n$.}

\mbox{}

\noindent {\bf Proof.} Consider the first $\rho$ rows of ${\cal J}$ (namely those comprising ${\cal J}^{[1]}$ and ${\cal J}^{[2]}$). Thus for $i=1, \ldots , \rho$ and for $l= \rho +1, \ldots , n$ we have:
\[
	({\cal J} \cdot \nabla D_l)_i = \sum_{j=1}^n J_{ij} \partial_j D_l =
	-  \sum_{j=1}^{\rho} J_{ij}^{[1]} a_{lj}+ \sum_{j= \rho +1}^{n} J_{ij}^{[2]} \delta_{lj} 
	= 0
\]
where $\delta_{lj}$ denotes Kronecker's delta and the last equality is a consequence of 
(\ref{n1j2}). Analogously, consider now the last $(n- \rho )$ rows of ${\cal J}$ (which involve submatrices ${\cal J}^{[3]}$ and ${\cal J}^{[4]}$). Now for $i,l= \rho +1, \ldots ,n$ we arrive at:
\[
	({\cal J} \cdot \nabla D_l)_i = \sum_{j=1}^n J_{ij} \partial_j D_l =
	-  \sum_{j=1}^{\rho} J_{ij}^{[3]} a_{lj}+ \sum_{j= \rho +1}^{n} J_{ij}^{[4]} \delta_{lj} 
	= 0
\]
where now the last identity arises as a result of (\ref{n1j43}). The proof Lemma 2 is complete. \hfill $\Box$

\mbox{}

\noindent {\bf Lemma 3.} {\em For all $i,j= 1, \ldots , n$, functions $J_{ij}(x)$ entering matrix ${\cal J}$ in (\ref{n1jlin}) are KG for ${\cal J}$.}

\mbox{}

\noindent {\bf Proof.} By construction in (\ref{n1jlin}), all entries $J_{ij}$ are either functions $\psi_{ij}(D_{\rho +1}(x), \ldots ,D_n(x))$, as it is the case of ${\cal J}^{[1]}$, or linear combinations with constant coefficients of such functions (as happens for ${\cal J}^{[2]}$, ${\cal J}^{[3]}$ and ${\cal J}^{[4]}$). Therefore, in a compact and unified way it can be said that there exist real constants $b_{ijkl}$ such that:
\begin{equation}
\label{n1l3d}
	J_{ij} = \sum_{\stackrel{\scriptstyle k,l=1}{\scriptstyle l > k } }^{\rho}
	 b_{ijkl} \psi_{kl}(D_{\rho +1}(x), \ldots ,D_n(x))
	\;\: , \;\:\;\: i,j =1, \ldots , n
\end{equation}
From (\ref{n1l3d}) it can be seen that:
\begin{equation}
\label{n1l3g}
	\nabla J_{ij} = \sum_{\stackrel{\scriptstyle k,l=1}{\scriptstyle l > k } }^{\rho}
	 b_{ijkl} \sum_{q= \rho +1}^n \left( \frac{\partial \psi_{kl}}{\partial D_q} \right) 
	\nabla D_q \;\: , \;\:\;\: i,j =1, \ldots , n
\end{equation}
Finally, as a consequence of (\ref{n1l3g}) and Lemma 2 we obtain ${\cal J} \cdot \nabla J_{ij}=0$ for $i,j=1, \ldots ,n$. Lemma 3 is thus shown. \hfill $\Box$

\mbox{}

Therefore it is proved that ${\cal J}$ is a D-solution. Then as a consequence of Lemma 2, the $(n- \rho)$ functions $D_l$ in (\ref{n1casl}) are Casimir invariants, for which the functional independence is clear. To complete the proof, note that by construction the rows $\rho +1, \ldots ,n$ of ${\cal J}$ are linear combinations (with constant coefficients) of the first $\rho$ ones, as implied by (\ref{n1j3}) and (\ref{n1j42}). Then it is clear that Rank(${\cal J}$) $\leq \rho$. Since the rank of a skew-symmetric matrix is always even, from Lemma 1 this means that necessarily it is Rank(${\cal J}$) $\leq \rho - \rho \bmod 2$, which is the bound given. This completes the proof of the theorem. \hfill $\Box$

\mbox{}

Of course, in the solution family described in Theorem 4 the linear dependences among the elements of ${\cal J}$ have been chosen in such a way that the $\rho$ first rows and columns  (those conforming ${\cal J}^{[1]}$) span the rest of rows and columns, as it is clear from equations (\ref{n1j2}--\ref{n1j43}). Actually, this choice is entirely arbitrary and was used without loss of generality. Analogous families of D-solutions can be generated on the basis of the rest of possible $\rho \times \rho$ submatrices. This fact, the convenience for future use and the notation employed in (\ref{n1jlin}) motivate the following definition:

\mbox{}

\noindent {\bf Definition 4.} {\em Every D-solution being either of the type constructed in Theorem 4 or a permutation of such construction will be termed a D$_{\psi}$-solution.}

\mbox{}

After presenting the main definitions and results, the purpose of the next section will be to provide some comments and examples aimed to clarify the content and implications of the previous developments. 

\mbox{}

\mbox{}

\begin{flushleft}
{\bf 3. Discussion and examples}
\end{flushleft}

The first basic aspect of D-solutions which is interesting to consider regards the possible rank values of D-solutions in general, and of D$_{\psi}$-solutions in particular, for a given value of $n$. The fact that D-solutions are composed by Casimir invariants seems to suggest that such Poisson structures must be degenerate and therefore that the maximal possible rank (namely $n - n \bmod 2$) should be excluded for D-solutions having an even value of $n$. However such statement is not mathematically correct, as the following example displays.

\mbox{}

\noindent {\bf Example 1.} Constant structure matrices play a significant role in the theory of Poisson systems. In first place, they comprise as a special case the classical symplectic matrices (and therefore the whole classical Hamiltonian theory \cite{z1gol} together with the canonical formalism). In addition, in the case of constant structure matrices different from the symplectic ones, they are the source of very varied noncanonical applications in different domains such as mechanics \cite{srw}, plasma physics \cite{25} or population dynamics 
\cite{ker}, just to cite a sample. Every constant structure matrix of arbitrary rank is a 
D-solution, since it is entirely composed by (trivial, namely constant) Casimir invariants: as said, constants are Casimir invariants for every structure matrix. This is interesting because constant structure matrices are able to take all possible ranks associated to every given value of $n$. Therefore it is worth emphasizing the next statement:

\mbox{}

\noindent {\bf Corollary 3.} {\em For every integer $n \geq 3$ and for every possible nontrivial rank value ($r$ even, $2 \leq r \leq n - n \bmod 2$) there exists an infinity of
$n$-dimensional D-solutions having constant rank $r$ in $\mathbb{R}^n$.}

\mbox{}

\noindent In particular, constant structure matrices also arise from Theorem 4 as 
D$_{\psi}$-solutions. To see this, it suffices to consider the value $\rho =n$. In such case, there are $(n- \rho)=0$ functions $D_l$ of the form (\ref{n1casl}) and ${\cal J}$ in 
(\ref{n1jlin}) is entirely given by submatrix ${\cal J}^{[1]}$, in other words ${\cal J}={\cal J}^{[1]}$. Moreover, functions $J_{ij}^{[1]}=\psi_{ij}(D_{\rho +1}(x), \ldots ,D_n(x))$ lose their dependences and become arbitrary constants. Consequently, ${\cal J}$ is an arbitrary constant skew-symmetric matrix, which can have every admissible (even) rank. Thus, in this situation we still have D$_{\psi}$-solutions, but in the limit case in which no Casimir invariants of the form (\ref{n1casl}) are implemented. 

Let us now present a first example of D$_{\psi}$-solution. Such instance has been chosen to illustrate the fact that the upper bound to the rank may not be reached.

\mbox{}

\noindent {\bf Example 2.} The following matrix corresponds to the form studied in Theorem 4 with $n=4$, $\rho =2$, and with the two linear Casimirs $D_3(x)=x_2+x_3$ and $D_4(x)=x_1+x_2+
x_4$ of the kind (\ref{n1casl}) implemented. If $\psi (y_1,y_2): \mathbb{R}^2 \rightarrow \mathbb{R}$ is an arbitrary $C^{\infty}(\mathbb{R}^2)$ function, the matrix is:
\begin{equation}
\label{n1jex2}
	{\cal J} = \left( \begin{array}{ccccc}
	 0 & \psi (D_3,D_4) & \vline & - \psi (D_3,D_4) & - \psi (D_3,D_4) \\
	- \psi (D_3,D_4)  &  0 & \vline & 0 & \psi (D_3,D_4) \\ \hline
	\psi (D_3,D_4) & 0 & \vline & 0 & - \psi (D_3,D_4)   \\ 
	\psi (D_3,D_4) & - \psi (D_3,D_4) & \vline & \psi (D_3,D_4) & 0 \\
	\end{array} \right) 
	\equiv \left( \begin{array}{ccc} 
		{\cal J}^{[1]} & \vline & {\cal J}^{[2]} \\ \hline
		{\cal J}^{[3]} & \vline & {\cal J}^{[4]}
	    \end{array} \right)
\end{equation}
It is evident that ${\cal J}$ in (\ref{n1jex2}) is skew-symmetric, that both $D_3(x)$ and $D_4(x)$ are KG functions for such matrix, and that all the entries of ${\cal J}$ are $C^{\infty}$ functions of $D_3(x)$ and $D_4(x)$. Moreover, the structure (\ref{n1jlin}) of the matrix (or, more in detail, the identities (\ref{n1j2}--\ref{n1j43}) among the four submatrices) can be easily verified. Accordingly, ${\cal J}$ in (\ref{n1jex2}) is a D$_{\psi}$-solution. In addition, for the rank we obtain from Theorem 4 the bound Rank(${\cal J}$) $\leq \rho - \rho \bmod 2 =2$, which will be in general the case almost everywhere in $\mathbb{R}^4$, but not in the hypersurface $\psi (D_3(x),D_4(x))=0$ where Rank(${\cal J}$) vanishes.

\mbox{}

Another aspect of interest of D$_{\psi}$-solutions is the possible presence of nonlinear Casimir invariants, additional to those implemented in (\ref{n1casl}). Associated to this issue is again the fact that Rank(${\cal J}$) can take lower values than the upper bound $\rho - \rho \bmod 2$. These possibilities are illustrated in the next instance.

\mbox{}

\noindent {\bf Example 3.} The following matrix is a D$_{\psi}$-solution, this time with $n=4$ and $\rho =3$:
\begin{equation}
\label{n1jex3}
	{\cal J} = \left( \begin{array}{ccccc}
		0      &   x_2+x_4   &  (x_2+x_4)^2 & \vline & -x_2-x_4  \\
	  -x_2-x_4   &      0      &       0      & \vline  &    0     \\
	-(x_2+x_4)^2 &      0      &       0      & \vline  &    0     \\ \hline
	   x_2+x_4   &      0      &       0      & \vline  &    0     \\
	\end{array} \right) 
	\equiv \left( \begin{array}{ccc} 
		{\cal J}^{[1]} & \vline & {\cal J}^{[2]} \\ \hline
		{\cal J}^{[3]} & \vline & {\cal J}^{[4]}
	    \end{array} \right)
\end{equation}
In this case, the linear function of the form (\ref{n1casl}) defined is $D_4(x)=x_2+x_4$. Thus ${\cal J}$ in (\ref{n1jex3}) is a D-solution because it is skew-symmetric and $D_4(x)$ is a KG function for ${\cal J}$. Actually, the submatrix structure (\ref{n1subj}) can be identified without difficulty for all ${\cal J}^{[i]}$, $i=1, \ldots ,4$, according to the constructive steps of the proof of Theorem 4. In addition, Theorem 4 predicts the bound Rank(${\cal J}$) $\leq \rho - \rho \bmod 2 =2$. Accordingly, at least one additional functionally independent Casimir invariant must exist. Actually, it can be easily verified that it is given by $D_3(x) = x_3+x_2x_4+x_4^2$, which is obviously nonlinear. In fact, both $D_3$ and $D_4$ are functionally independent as we can see on their Jacobian:
\begin{equation}
\label{n1jace3}
	\frac{\partial (D_3,D_4)}{\partial (x_1,x_2,x_3,x_4)}=
	\left( \begin{array}{cccc}
		0 & x_4 & 1 & x_2+2x_4 \\
		0 &  1  & 0 & 1
	\end{array} \right)
\end{equation}
Functional independence is thus proven since Jacobian (\ref{n1jace3}) has rank 2 everywhere in $\mathbb{R}^4$. On the other hand, this set of Casimir invariants is not complete when the upper bound Rank(${\cal J}$) $= 2$ is not accomplished. Actually this is possible, since all the entries of ${\cal J}$ vanish in the hyperplane $x_2+x_4=0$. On the contrary, Rank(${\cal J}$) $= 2$ everywhere else in $\mathbb{R}^4$, thus complying to the upper bound predicted by Theorem 4. In such case, Casimir invariants $D_3$ and $D_4$ constitute a complete set.

\mbox{}

A relevant question indirectly posed by Theorem 4 regards the construction of D-solutions for which the implemented Casimirs are nonlinear. In general, the method should consist of the {\em a priori\/} specification of $(n- \rho )$ future independent Casimir invariants of the form
\begin{equation}
\label{n1cnl}
	D_i(x)= x_i - \mu_i(x_1, \ldots ,x_{\rho}) \:\; , \:\;\:\; i= \rho +1, \ldots ,n
\end{equation}
where the $\mu_i(x_1, \ldots ,x_{\rho})$ are smooth functions. Let $R_i$ denote the $i$-th row of the matrix. According to the procedure used in Theorem 4 for the case of linear invariants, we have that $R_i = \sum_{j=1}^{\rho}(\partial _j \mu_i)R_j$ for $i= \rho +1, \ldots ,n$ in the matrix to be obtained, together with a similar relationship for the columns, so that the $D_i(x)$ in (\ref{n1cnl}) are KG functions by construction. Similarly to what is done in Theorem 4, ${\cal J}$ should be splitted in four regions, with ${\cal J}^{[1]}$ defined as in (\ref{n1jlin}), so that the entries of ${\cal J}^{[1]}$ are KG functions of the resulting matrix. This is exactly the procedure used in Theorem 4, but applied to the nonlinear functions 
(\ref{n1cnl}). The problem however is that the outcome for ${\cal J}^{[2]}$, ${\cal J}^{[3]}$ and ${\cal J}^{[4]}$ generally produces functions that are not KG functions of the matrix thus constructed. The reason for this is that now the coefficients of the linear combinations generating the entries of ${\cal J}^{[2]}$, ${\cal J}^{[3]}$ and ${\cal J}^{[4]}$ are of the form $\partial _j \mu_i$, namely they are not constant (as in the case of Theorem 4). The outcome is that the entries of ${\cal J}^{[2]}$, ${\cal J}^{[3]}$ and ${\cal J}^{[4]}$ lose in general their functional dependence with respect to functions (\ref{n1cnl}). On the contrary, such functional dependence is preserved in the case of linear invariants (\ref{n1casl}) because (in the notation of Theorem 4) in this situation we have that $\partial _j \mu_i = a_{ij}$ is always a constant. A simple example now illustrates this situation.

\mbox{}

\noindent {\bf Example 4.} Let us consider the case $n=3$, $\rho=2$ for the implementation of a nonlinear Casimir invariant of the form $D_3(x)=x_3- \mu (x_1,x_2)$. For $D_3(x)$ to be by construction a KG function, it has to be $R_3 = (\partial _1 \mu) R_1 + (\partial _2 \mu) R_2$, and a similar relationship for the columns. Thus defining $J_{12}(x)= \psi (D_3(x))$ we are led to the matrix:
\begin{equation}
\label{n1e4m}
	{\cal J}= \psi (D_3(x))
	\left( \begin{array}{ccc}
	0 & 1 & \partial_2 \mu (x_1,x_2) \\
	- 1 &  0  & - \partial_1 \mu (x_1,x_2) \\ 
	- \partial_2 \mu (x_1,x_2) & \partial_1 \mu (x_1,x_2) & 0 
	\end{array} \right)
\end{equation}
Now it is evident that ${\cal J}$ in (\ref{n1e4m}) is skew-symmetric and that both $D_3(x)$ and $J_{12}(x)$ are KG functions for ${\cal J}$, as expected. However, it is straightforward to check that the rest of nondiagonal entries are not in general KG functions of ${\cal J}$. Consequently, ${\cal J}$ in (\ref{n1e4m}) is generally not a D-solution. Notice that in the particular case (\ref{n1casl}) of linear Casimir functions the partial derivatives of $\mu (x_1,x_2)$ are constant as indicated, and then every entry of ${\cal J}$ in (\ref{n1e4m}) is a KG function, thus conforming a D$_{\psi}$-solution. On the contrary, when functions $\partial_i \mu (x_1,x_2)$ are not constant, the property of being a D-solution is not necessarily preserved.

\mbox{}

Since the previous reasoning is clear, the naturalness of the result contained in Theorem 4 suggests yet another relevant question, namely the possibility that all the solutions of the distinguished Jacobi equations (\ref{n1jsksym},\ref{n1rjac}) are actually D$_{\psi}$-solutions. This is certainly not the case, as it can be seen in what follows.

\mbox{}

\noindent {\bf Example 5.} In this example D-solutions of arbitrary dimension $n \geq 3$ and not being D$_{\psi}$-solutions will be considered. In fact, such D-solutions do not have linear Casimir invariants at all, as we shall see in what follows. Consider the following $n \times 
n$ skew-symmetric matrix:
\begin{equation}
\label{n1e5mn}
  {\cal J}= 
  \left( \begin{array}{cccccc}
  0 & \vline & x_3/x_2 & (x_3/x_2)^2 & \ldots & (x_3/x_2)^{n-1} \\ \hline 
  - x_3/x_2 & \vline &  \mbox{} & \mbox{} & \mbox{} & \mbox{} \\ 
  -(x_3/x_2)^2 & \vline &  \mbox{} & \mbox{} & \mbox{} & \mbox{} \\ 
   \vdots & \vline & \mbox{} & \mbox{} & O_{(n-1) \times (n-1)} & \mbox{} \\
  -(x_3/x_2)^{n-1} & \vline &  \mbox{} & \mbox{} & \mbox{} & \mbox{} 
	\end{array} \right)
\end{equation}
In (\ref{n1e5mn}) the symbol $O$ denotes a null submatrix of size indicated by the respective subindex. This matrix will be defined in a domain $\Omega \subset \mathbb{R}^n$ in which $x_2 \neq 0$ and $x_3 \neq 0$ for every $x \in \Omega$. In such case, we have Rank(${\cal J}$) $=2$ everywhere in $\Omega$. It is then possible to find $(n-2)$ KG functions for matrix 
(\ref{n1e5mn}) functionally independent in $\Omega$. These are:
\begin{equation}
\label{n1cen}
	D_3(x)= \frac{x_3}{x_2} \:\; ; \:\; \:\; 
	D_i(x) = \frac{x_3x_{i-1}}{x_2}-x_i \:\; , \:\; \:\; i = 4, \ldots ,n
\end{equation}
That they are KG functions is simple to verify. Regarding functional independence, notice that we have the Jacobian:
\begin{equation}
\label{n1jen}
	\left( \frac{\partial (D_3, D_4, \ldots ,D_n)}{\partial (x_1, \ldots ,x_n)} \right)^T =
	\left( \begin{array}{cccc}
		0 & 0 & \ldots & 0 \\
		-x_3/x_2^2 & -x_3^2/x_2^2 & \ldots & -x_3x_{n-1}/x_2^2 \\
		1/x_2 & 2x_3/x_2 & \ldots & x_{n-1}/x_2 \\
		0 & -1 & \ldots & 0 \\
		0 & 0 & \ldots & 0 \\
		\vdots & \vdots & \mbox{} & \vdots \\
		0 & 0 & \ldots & x_3/x_2 \\
		0 & 0 & \ldots & -1 
	\end{array} \right)
\end{equation}
Thus if we choose in (\ref{n1jen}) the submatrix composed by the last $(n-2)$ rows, we see that it is upper triangular with determinant $(-1)^{n-3}(x_2)^{-1} \neq 0$ in $\Omega$. Accordingly, functional independence of $D_3(x), \ldots ,D_n(x)$ holds in $\Omega$ for every $n \geq 3$. Since all the entries of ${\cal J}$ in (\ref{n1e5mn}) are $C^{\infty}$ functions of $D_3(x)$ we have that such matrix is a D-solution for every $n \geq 3$, and functions $D_3(x), \ldots ,D_n(x)$ in (\ref{n1cen}) form a complete set of independent Casimir invariants of ${\cal J}$ in 
$\Omega$. On the other hand, recall that if a D$_{\psi}$-solution of constant rank has one or more independent Casimir invariants, then at least one of them can be taken to be linear: in the case $\rho <n$ this is so by construction; and according to Example 1, in the complementary case $\rho =n$ the matrix is constant, which implies the existence of linear Casimir invariants when such matrix is degenerate. Conversely, a degenerate structure matrix of constant rank without linear Casimir invariants is not a D$_{\psi}$-solution. This is evidently the case for matrix ${\cal J}$ in (\ref{n1e5mn}) since the {\em ansatz\/} of a generic linear Casimir invariant $D(x)= \sum_{i=1}^n a_ix_i$ substituted in the identity ${\cal J} \cdot \nabla D =0$ immediately leads to $a_i=0$ for all $i=1, \ldots ,n$. Therefore it is proved that ${\cal J}$ in (\ref{n1e5mn}) is a D-solution but not a D$_{\psi}$-solution, for every $n \geq 3$.

\mbox{}

Note in addition that the significance of the last example is reinforced in view of the results displayed in Theorems 1, 2 and 3. This motivates the following conclusion:

\mbox{}

\noindent {\bf Corollary 4.} {\em For every $n \geq 3$, there exists an infinity of 
$n$-dimensional D-solutions that are not D$_{\psi}$-solutions.}

\mbox{}

In other words, D$_{ \psi }$-solutions do not provide the general solution of the distinguished Jacobi equations (\ref{n1jsksym},\ref{n1rjac}). In the next section we explore further these issues by considering the properties of three-dimensional D$_{\psi}$-solutions and some of their applied consequences. 

\mbox{}

\mbox{}

\begin{flushleft}
{\bf 4. Physical interpretation of three-dimensional D$_{\psi}$-solutions and application to their global reduction to the Darboux canonical form}
\end{flushleft}

In what follows we consider the case of three dimensions, in which it is possible to further investigate the families of D-solutions and D$_{\psi}$-solutions. From a physical point of view, we shall see that in this case the D$_{\psi}$-solution condition points out the significance of certain time reparametrizations for Poisson systems. As a consequence, it is shown that the Darboux canonical form can be globally constructed for such family of Poisson structures.

\mbox{}

\noindent {\bf Theorem 5.} {\em Let ${\cal J} \equiv (J_{ij})$ be a $3 \times 3$ matrix defined in a domain $\Omega \subset \mathbb{R}^3$ and having the form:  
\begin{equation}
\label{3ddsg}
	{\cal J}= \eta(a_1x_1+a_2x_2+a_3x_3) \left( \begin{array}{ccc} 
	0 & a_3 & -a_2 \\ -a_3 & 0 & a_1 \\ a_2 & -a_1 & 0 \end{array} \right)
\end{equation}
where $a_1$, $a_2$, $a_3$ are real constants, and $\eta(y)$ is a single-variable 
$C^{\infty}(\mathbb{R})$ function. Then:
\begin{itemize}
\item[\mbox{\rm \mbox{\em (a)}}] Every matrix of the form (\ref{3ddsg}) is a 
D-solution in $\Omega$.
\item[\mbox{\rm \mbox{\em (b)}}] A three-dimensional structure matrix defined in $\Omega$ is a D$_{\psi}$-solution if and only if it has the form (\ref{3ddsg}). 
\item[\mbox{\rm \mbox{\em (c)}}] If $a_i \neq 0$ at least for one value of $i=1,2,3$ and 
$\eta(a_1x_1+a_2x_2+a_3x_3) \neq 0$ for every $(x_1,x_2,x_3) \in \Omega$, then $D(x_1,x_2,x_3)=a_1x_1+a_2x_2+a_3x_3$ is the only independent Casimir invariant of ${\cal J}$ in $\Omega$, and 
${\cal J}$ can be globally and constructively reduced in $\Omega$ to the Darboux canonical form. 
\end{itemize}
}

\mbox{}

\noindent {\bf Proof.} (a) Let us first recall that every constant skew-symmetric matrix is a structure matrix (see the discussion in Example 1). Consequently, the following matrix ${\cal J}_0$ is a D-solution:
\[
	{\cal J}_0 = \left( \begin{array}{ccc} 
	0 & a_3 & -a_2 \\ -a_3 & 0 & a_1 \\ a_2 & -a_1 & 0 \end{array} \right)
\]
Now it is direct to verify that function $D(x_1,x_2,x_3)=a_1x_1+a_2x_2+a_3x_3$ is a Casimir invariant of ${\cal J}_0$ globally defined in $\Omega$ since ${\cal J}_0 \cdot \nabla D =0$. Then the result arises after the application of Theorem 1.e. 

\mbox{}

(b) According to Theorem 4, for $n=3$ we have the possible values of $\rho =1,2,3$. For the cases $\rho =1$ and $\rho = 3$ we obtain constant matrices, including the case of zero rank. Let us focus on the value $\rho =2$. From (\ref{n1casl}) we have $D_3(x_1,x_2,x_3)=x_3-a_{31}x_1-a_{32}x_2$, and then from 
(\ref{n1jlin}) we obtain $J_{12}= \psi_{12}(D_3)$, $J_{12}= a_{32} \psi_{12}(D_3)$, and 
$J_{23}= -a_{31}\psi_{12}(D_3)$. It is direct to see that this is equivalent to (\ref{3ddsg}) in the case $a_3 \neq 0$. Taking into account Definition 4, we conclude that similar constructions can be developed for the complementary cases $a_1 \neq 0$ and $a_2 \neq 0$. Moreover, the zero-rank case is also included in the construction of D$_{\psi}$-solutions, as indicated. Therefore, all three-dimensional D$_{\psi}$-solutions are given by expression (\ref{3ddsg}). 

\mbox{}

(c) The hypotheses assumed imply that Rank(${\cal J}$) $=2$ everywhere in $\Omega$. Let us recall that $D$ is a Casimir invariant of ${\cal J}$. Since Rank(${\cal J}$) $=2$ everywhere in $\Omega$, the Casimir $D$ is the only one functionally independent in $\Omega$. Moreover, constancy of the rank is a necessary condition for the existence of the Darboux canonical form \cite{olv1}. Accordingly, the algorithm developed in Theorem 5.1 of reference \cite{bnn2} can be applied in order to construct globally in $\Omega$ the Darboux canonical form for every structure matrix of the form (\ref{3ddsg}). 

This completes the proof of the theorem. \hfill $\Box$

\mbox{}

For the sake of illustration, it can be mentioned that some D-solutions of the form 
(\ref{3ddsg}) can be found in \cite{creh}. In Theorem 5 we have considered the three-dimensional case in order to gain additional insight on the meaning of D-solutions in the framework provided by the general Poisson structures. As it can be seen from Theorem 5, together with Theorem 1.e, we have that three-dimensional D$_\psi$-solutions consist essentially of a multiplicative deformation of a constant Poisson structure. In fact, such kind of deformations (based on the multiplication by any Casimir invariant) are closely related to the use of time reparametrizations preserving the Poisson structure \cite{bnn2,bnn4}. This provides a physical interpretation of D$_{\psi}$-solutions and of the restriction imposed by equations (\ref{n1jsksym},\ref{n1rjac}), at least in the present 3-d context. As we have seen in the last proof, the dynamical relationship with the use of time-reparametrizations is an important issue that allows the global and constructive determination of the Darboux canonical form for family (\ref{3ddsg}). In this sense, the reader is referred to Theorem 5.1 of reference \cite{bnn2} for full details. In addition it is worth recalling that, in principle, the Darboux construction only exists locally \cite{olv1}, and even this local construction may be a complicated task in general. When determined globally, the Darboux canonical form provides an application that leads to the establishment of a fundamental physical and dynamical link between Poisson and classical Hamiltonian systems 
\cite{7},\cite{bs2}-\cite{bnn3},\cite{bnnm,byv2,byv4,25}. To complete the discussion, it is interesting to mention also that the result provided in Theorem 5.b is complemented by Corollary 4, in the sense that the set of D$_{\psi}$-solutions is not equivalent to (but only a proper subset of) the general family of D-solutions, also in the three-dimensional case. 

In the next section we conclude by briefly regarding some of the previous issues as well as other questions from a more general perspective.

\mbox{}

\mbox{}

\begin{flushleft}
{\bf 5. Final remarks}
\end{flushleft}

The study of solutions of the Jacobi equations provides an increasingly rich perspective of finite-dimensional Poisson structures. In spite that a complete knowledge of such solutions is still far, the investigation of the problem seems to be not only a mathematically appealing subject, but also a unavoidable issue for a better understanding of finite-dimensional Poisson systems, and therefore of the scope of Hamiltonian dynamics. 

As discussed in the Introduction, the Jacobi equations become increasingly complex as dimension grows. This explains that the characterization of families of arbitrary dimension composed by generic functions (namely not limited to a given degree of nonlinearity) and having arbitrary rank is still very uncommon in the literature (instances of the same kind are provided in 
\cite{bs07,bnnm,byv4}). For this reason, D-solutions may well be regarded as a significant contribution in such sense. 

When compared to the analyses of other solution families \cite{7},\cite{bs2}-\cite{bnn3},\cite{bnnm,byv2,byv4,25}, it can be seen that a typical outcome is the global construction of the Darboux canonical form. In principle, this is not possible here, neither in general nor in the specific case of D$_{\psi}$-solutions. This is mainly due to (a) the absence of a closed-form expression for D-solutions; (b) the generality of the functional form of the entries of the solution matrices ${\cal J}$; and (c) the possible lack of constancy of the rank (already illustrated in the examples). Together, these three items seem to exclude a global application of Darboux theorem in the case of D-solutions, at least with full generality. Nevertheless, it is possible in some relatively general situations, as shown in Section 4 for the three-dimensional family of D$_\psi$-solutions. 

From an interpretative point of view, at first sight it could be said that the existence of 
D-solutions means that for any Poisson matrix it is possible to replace numerical constants 
with Casimir invariants without changing the property of being a structure matrix. The basis for this incorrect replacement is that Casimir invariants are constants of motion of the associated Poisson system, and are independent of the Hamiltonian function, namely they are first integrals arising from the structure matrix, and thus they are functions having a constant value during the time evolution of the system. However, Casimir invariants are functions of the system variables and the replacement of one of such functions by a numerical constant, or vice versa, does not ensure the verification of the Jacobi equations (\ref{n1jjac}-\ref{n1jsksym}) for the resulting skew-symmetric matrix. Note that a Casimir invariant takes a constant value on a level set of the symplectic foliation, but the value of such constant is different for different points of phase-space. The fact that the Casimir invariants are first integrals of any Poisson system having such structure matrix does not suppress the functional form (generally non-constant) of such invariants. Actually, in the Jacobi equations (\ref{n1jjac}-\ref{n1jsksym}) the time variable does not appear, and therefore the time constancy of Casimir functions is not present in the problem. 

To conclude, notice also that in spite of the mathematically nice specialization of the general 
Jacobi equations which is provided by the D-solution problem, this does not seem to imply that even a complete identification of D-solutions is at hand, as Corollary 4 points out. Consequently, either in the general form or in the more specific distinguished version, it turns out to be natural that Jacobi equations will be the subject of further research in the future. 

\pagebreak

\end{document}